\documentstyle[aps,twocolumn,floats,prl,epsfig]{revtex}
\begin{document}
\title{Ground state spin and Coulomb blockade peak motion in chaotic 
quantum dots}
\author{J.~A.~Folk$^{1,2}$, C.~M.~Marcus$^1$, R.
Berkovits$^{3,4,5}$, I.~L~.Kurland$^{3}$, I.~L.~Aleiner$^6$,
B.~L.~Altshuler$^{3,4}$}
\address{$^{1}$ Department of Physics, Harvard University, Cambridge, MA 02138}
\address{$^{2}$ Department of Physics, Stanford University, Stanford, CA 94305}
\address{$^{3}$ Physics Department, Princeton University, Princeton, NJ 08544}
\address{$^{4}$ NEC Research Institute, 4 Independence Way, 
Princeton, NJ 08540}
\address{$^{5}$ Minerva Center and Department of Physics, Bar-Ilan 
University, Ramat-Gan 52900,
Israel}
\address{$^{6}$ Department of Physics, SUNY at Stony Brook, Stony Brook, NY 11794
}

\draft
\maketitle
\begin{abstract}  We investigate experimentally and theoretically the 
behavior of Coulomb blockade (CB) peaks in
a magnetic field that couples principally to the ground-state spin 
(rather than the orbital moment)
of a chaotic quantum dot. In the first part, we discuss  numerically 
observed features in the magnetic field dependence of CB peak and spacings that
unambiguously identify changes in spin
$S$ of each ground state for successive numbers of electrons on the 
dot, $N$.  We next evaluate the
probability that the ground state of the dot has a particular spin 
$S$, as a function of the exchange strength,
$J$, and external magnetic field, $B$.  In the second part, we 
describe recent experiments on
gate-defined GaAs quantum dots in which Coulomb peak motion and 
spacing are measured as a function of
in-plane magnetic field, allowing changes in spin between $N$ and 
$N+1$ electron ground states to be
inferred.
\end{abstract}

\section{Introduction}
\label{s1}

In the absence of interactions, electrons populate the orbital 
states of a quantum dot or
metallic grain in an alternating sequence of spin-up and spin-down 
states, in accordance with the
Pauli picture. In this case, the total spin of the dot is either 
zero, when the number of
electrons $N$ is even, or one half at odd $N$. No higher spin states 
 appear. It is well known
that in the presence of interactions, this simple scheme can be 
violated. Such deviations from
simple even-odd filling may appear as Hund's rules in atomic physics, 
or be as extreme as complete
spin polarization leading to ferromagnetism by the mechanism of the 
Stoner instability.

  By measuring electron transport through a weakly coupled quantum dot 
at low temperature and bias
voltage (compared to the quantum level spacing) one can effectively 
study differences in ground state
(GS) properties of the dot at successive electron number, providing a 
means of investigating the
actual filling scheme
\cite{dotreview}.  A standard experimental approach
\cite{dotreview} is to measure conduction through the dot via two 
tunneling leads as a function of a
voltage,
$V_g$, applied between the dot and  a third electrically isolated 
electrode, known as a gate.
Because the dot GS energy, $E_N$, at a fixed electron
number is influenced by $V_g$, the gate can be used to set the number of 
electrons on the dot.
  The fact that a large Coulomb energy is needed to add a single 
electron to the dot typically
suppresses conduction through the dot in the tunneling regime, where 
whole charges must tunnel for
conduction to occur. This effect is known as the Coulomb blockade 
(CB).  However, at specific values
of gate voltage, denoted
$V_g^{(N)}$, where the condition
$E_{N-1}(V_g)= E_{N}(V_g)$ is satisfied, conductance increases 
dramatically, clearly marking this
degeneracy condition in an experimental trace. The position, $V_{g}^{(N)}$, of the
$N^{th}$ peak
  in the conductance is proportional to
$\mu_N = E_{N}(0)-E_{N-1}(0)$. Accordingly, the distance between 
successive peaks is $\Delta_2^{N} = \mu_N -
\mu_{N-1}$. (We take the constant of proportionality converting gate 
voltage to dot energy to be unity for the theoretical
discussion; experimentally, this constant can be readily measured, 
for instance, by comparing
the influence of bias and gate voltages.)

In the last few years the GS spin of a variety of nanostructures, including
  metallic grains\cite{ralph95}, semiconducting quantum dots
\cite{rokhinson00,duncan00,folk_opendot00,folk00}, and carbon nanotubes
\cite{cobden98,tans98}, have been investigated using CB peak motion 
in a magnetic field. If one
neglects the magnetic field coupling to the orbital degrees of 
freedom, one expects the field to
manifest itself only through the Zeeman splitting, resulting in a 
shift of  the GS energy by $g
\mu_B S B$,  where $\mu_B$ is the Bohr magneton  and $S$  denotes the 
GS total spin. For 2D quantum
dots, orbital coupling can be strongly suppressed---though not 
eliminated entirely---by orienting
the field strictly in the 2D plane.  This is the experimental 
approach that will be described in Section III
of this paper. On the other hand,  for ultrasmall grains and 
nanotubes, it is reasonable to ignore
orbital coupling for any field direction since for practical magnetic 
fields the total flux cutting
the structure is much less than the quantum of flux.

\begin{figure}[bth]\centering
\includegraphics[width=2.8in]{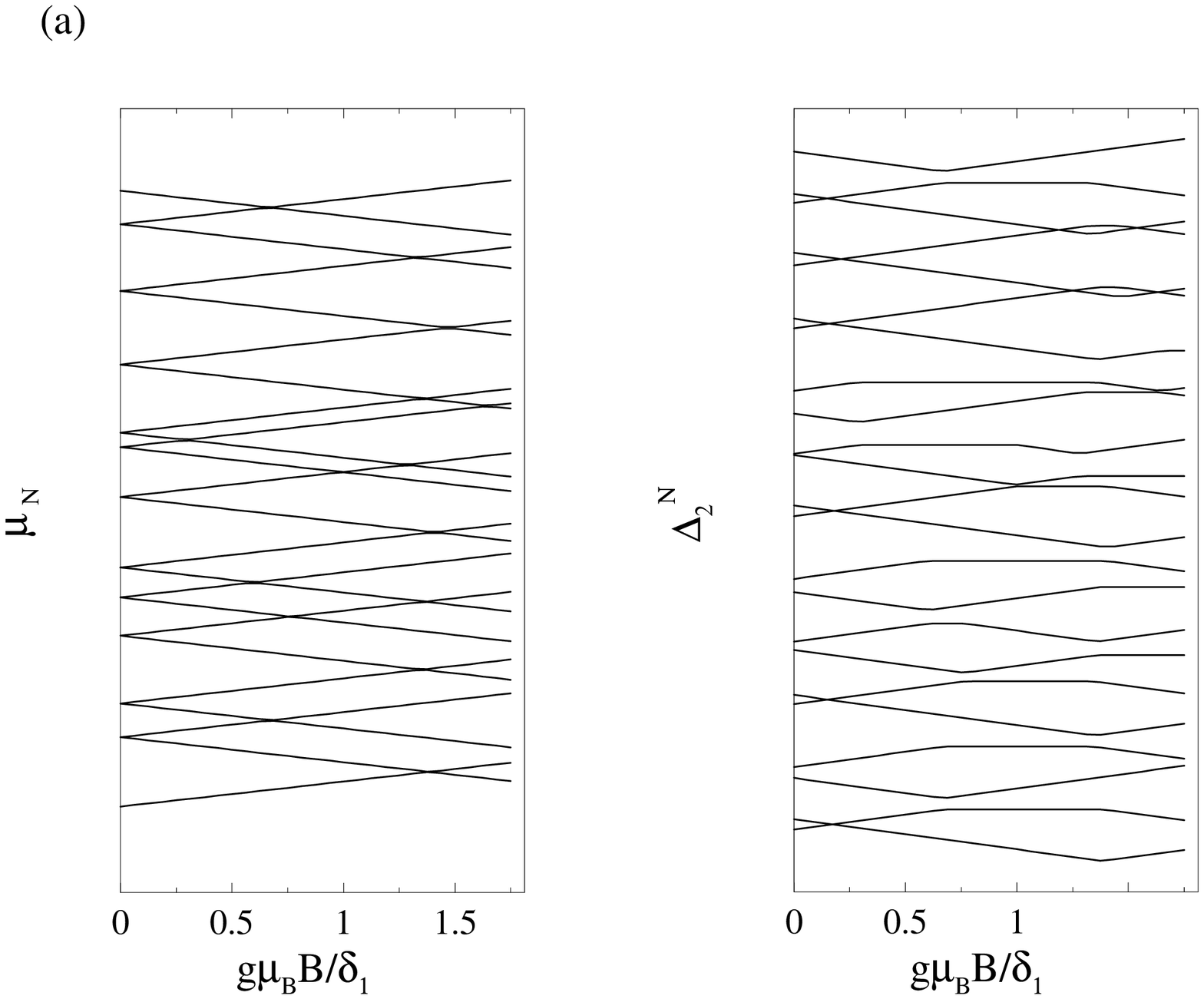}
\includegraphics[width=2.8in]{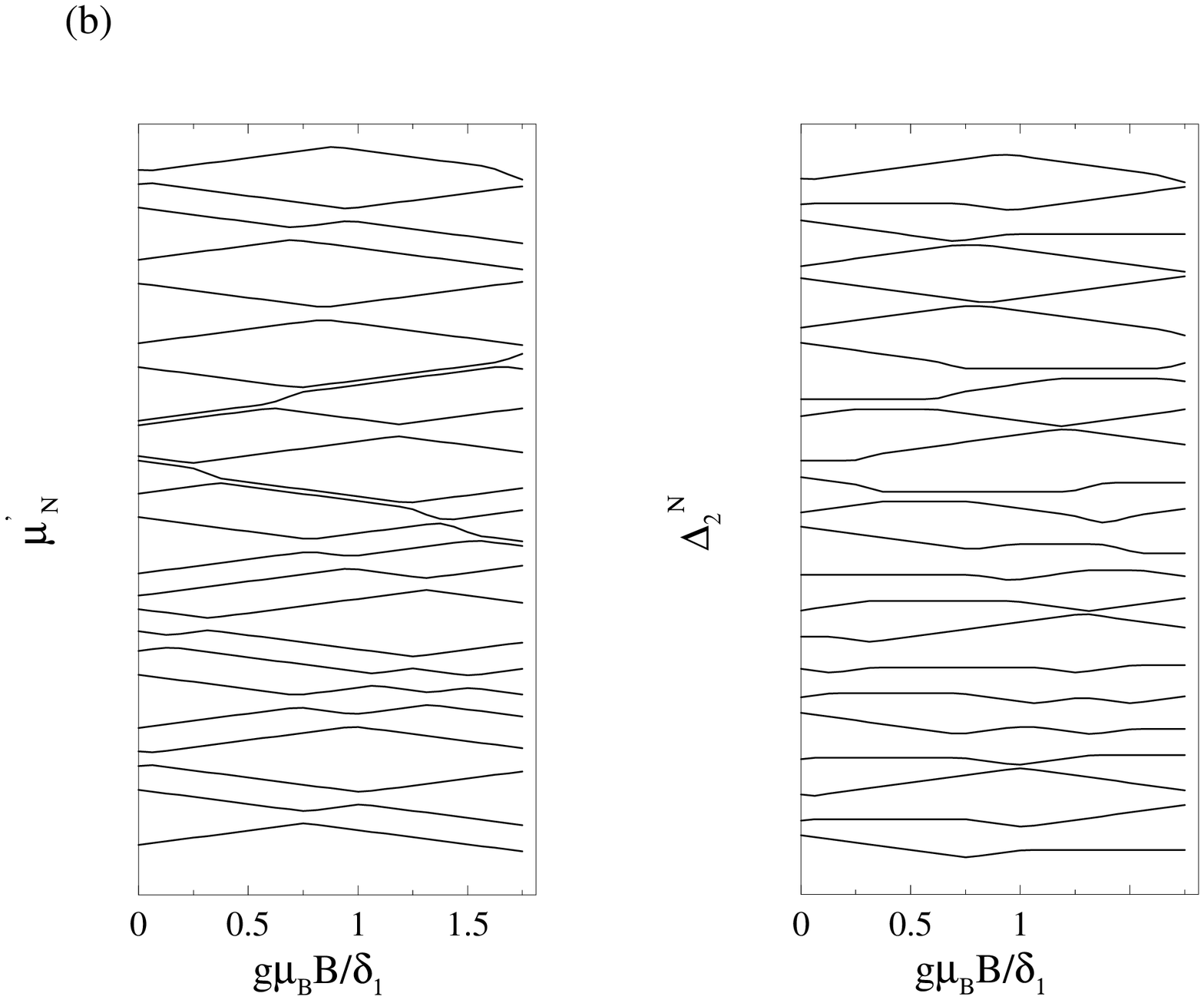}
 {\bigskip\caption[1cm]{\small  Typical 
conductance dependence on the gate voltage and
magnetic field, $B$, with and without exchange interaction: (a) 
$J=0$,  (b) $J=0.1 \delta_1$.
$\mu'_{N}$ is determined as
$\mu'_{N}=\mu_{N}-e^2/C$ and taken in arbitrary units. $\Delta_2^{N}$ 
for each $N$ is shifted
arbitrarly by $N \times$constant to avoid overlapping.}
\label{fig1}}\end{figure}

For  semiconducting quantum dots 
\cite{rokhinson00,duncan00,folk_opendot00,folk00} the Zeeman 
splitting for $S=1/2$
becomes comparable to the mean  single electron level spacing
\begin{equation}
\delta_1 = \langle s_i\rangle; \ \ \  s_i = \varepsilon_{i} - 
\varepsilon_{i+1},
\label{spacing}
\end{equation} (here $\varepsilon_{i}$ denotes the orbital energy of 
the one-electron  orbital state
$i$, and $\langle\ldots\rangle$ stands for an average over different 
levels) at $B \sim 1\,T$ for the GaAs dots
is Refs.\  \cite{duncan00,folk00} and
$B
\sim 10\,T$ for the Si dot in Ref.\ \cite{rokhinson00}.  At lower 
fields, the splitting seldom exceeds
$s_i$, and
  one might expect a simple even/odd Pauli picture. In this case, CB 
peak positions  would move with
$B$ as consecutive pairs, creating a pattern  of alternating downward 
and upward moving peaks, as
seen in Fig.\
\ref{fig1}a.  The trajectories of the pair of peaks would form 
straight lines with the same slope of $g \mu_b
/ 2$.  The peak {\em spacings},
$\Delta^N_2$, will have a similar pattern with slopes of $g \mu_b$. 
Spin-orbit interaction can lead
to fluctuations in the $g$ factor, resulting in fluctuations in the 
slopes of the lines,
\cite{matveev00,brouwer00}, but will not change the pattern of 
downward and upward moving peaks. Once
$g \mu_b B$ exceeds a particular spacing $s_i$, the peaks should cross.

Experimentally, the Pauli picture appears to work well for metallic 
grains \cite{ralph95}, while for
carbon nanotubes the situation is less clear: one published 
experiment finds simple even/odd filling
\cite{cobden98}, and one finds a more complicated scheme 
\cite{tans98}. For multielectron
semiconducting dots
\cite{rokhinson00,duncan00,folk00} the data are somewhat confusing, 
but appear to disagree with the
simple even/odd filling scheme
\cite{rokhinson00,duncan00,folk00,stewart98}, presumably as a result 
of electron-electron
interactions. In symmetric, few-electron vertical dots 
\cite{tarucha96} one may again recover
relatively simple behavior for the first few electrons, with a 
well-understood appearance of Hund's
rules. These conclusions are based on experiments where orbital 
magnetism dominates spin. However,
for few electrons (i.e., $N\lesssim 10$) the filling scheme can be 
readily interpreted, allowing the
full spin structure to be mapped out as a function of
$N$ and $B$
\cite{tarucha96}.

  Figure \ref{fig1} shows CB peak motion and spacing as a function of 
magnetic field (coupling only
to spin) for both a noninteracting electron model and an interacting 
electron model presented in detail in Section II. The  behavior
of this model appears more similar to the experimentally observed 
behavior in semiconducting quantum dots (see
Fig.\ \ref{fig6}).  In some cases peak positions ($\mu_N$) move as 
function of the magnetic field in
bunches of two  in the same direction and with the same slope, while 
their corresponding spacings
($\Delta_2^N$) do not change as function of the magnetic field. 
Moreover,  not every change in the
direction of the motion of a peak can be explained as a  crossing of 
two orbital levels.  Thus, a
description of the system beyond the Pauli picture is needed.

We will present a theory that goes beyond the simple Pauli picture 
and explains the puzzling features
of the magnetic field  dependence of the conductance peaks positions 
as a result of spontaneous spin
polarization  of the electrons in the dot \cite{kurland00}.   The 
possibility of  spontaneous
magnetization in quantum dots has  been considered in different contexts
\cite 
{kurland00,andreev98,berkovits98,brouwer99,eisenberg99,baranger00,jacquod00}.
For instance, it has also been suggested as an explanation
\cite{berkovits98}  for the absence of a  bimodal distribution in the 
conductance peak  spacings
\cite{exp_spacing}. Such a magnetization was also predicted to lead 
to kinks in the parametric
motion of the peaks,  (e.g., due to orbital effects of a 
perpendicular magnetic field)
\cite{baranger00}.

\section{Theory}
\label{s2}

We begin the theoretical discussion by reviewing the considerations 
determining the general form of
the Hamiltonian describing the properties of a chaotic dot in the 
metallic regime. A more complete
description may be found in Ref. \cite{kurland00}. The single 
electron spectrum $\varepsilon_{i}$ is
characterized by the mean level spacing $\delta_{1}$ and the Thouless 
energy $E_T \approx
\hbar/t_{erg}$, where $t_{erg}$ is the time it takes for a classical 
electron to cover the energy
shell in the single-particle phase space.  For diffusive and 
ballistic systems $t_{erg}$ equals to
$L^{2}/D$ and $L/v_F$ respectively, 
where $v_F$ is the electronic Fermi velocity, 
$L$ is the system size,
and $D$ is the diffusion coefficient. 
The dimensionless conductance, 
$g_T = E_{T}/\delta_{1}$, in the metallic
regime is large, i.e., $g_T \gg 1$. In this regime the statistics of 
the single electron spectrum on
scales smaller than
$E_{T}$ are well described by random matrix theory (RMT) 
\cite{mehta91}, which gives a quantitative
description of the phenomenon of level repulsion.  For an ensemble of
$M \times M$ matrices ($M \rightarrow \infty$)  with random and 
independent elements, the
probability density of a realization of the spectrum
$\varepsilon_{i}$ is given by\cite{mehta91}:
\begin{equation} P(\varepsilon_i) \propto \exp \left[\frac{\beta}{2} 
\sum_{i \neq j}\ln
\left(\frac  {\vert \varepsilon_i - \varepsilon_j \vert} 
{\delta_{1}}\right)\right],
\label{repl}
\end{equation}   where $\beta$ is equal to $1$, $2$ or $4$ for the 
orthogonal, unitary and symplectic
ensembles respectively.  The orthogonal (unitary) RM ensemble 
corresponds to weakly disordered dots
with preserved (violated) time reversal symmetry and negligible 
spin-orbit.    We assume that the dot is small enough for the spin-orbit
interaction to be negligible and thus avoid discussion of the symplectic
ensemble ($\beta=4$) \cite{halperin00}.

The second step is to consider effects of electron-electron 
interactions.   For the simplest case of
short-range interactions, we write
\begin{equation} H_{int}(\vec{r})=\lambda \delta_1 V \delta(\vec{r}),
\label{v.int}
\end{equation}  where $V\propto L^d$ is the volume of the dot (in $d$ 
dimensions) and
$\lambda$ is a dimensionless coupling constant characterizing the 
strength of the interaction. Matrix
elements of this interaction, in the basis of eigenstates
$\varphi_i(\vec{r})$ of the noninteracting Hamiltonian, are given by
\begin{equation} M^{i j}_{k l}=\lambda\delta_1 V\int d\vec{r}\varphi_{i}^{\ast}
(\vec{r})\varphi_{j}^{\ast}(\vec{r})\varphi_{k}(\vec{r})\varphi_{l}(\vec{r}).
\label{mtx1}
\end{equation}  It is important to note that the statistical 
properties of the interaction matrix
elements $M^{i j}_{k l}$ are completely determined by the statistical 
properties of the single
electron eigenstates $\varphi_i(\vec{r})$ and cannot be chosen in an 
arbitrary fashion as in the
random interaction model\cite{jacquod00}. In fact, in the limit $g_T \to
 \infty$ the diagonal matrix elements ($i,j,k,l$ pairwise equal)
are self-averaged by the space integration, Eq.~(\ref{mtx1}), and thus
show no level-to-level or sample-to-sample
fluctuations\cite{kurland00}.   In the same limit the off-diagonal matrix
elements Eq.~(\ref{v.int}) turn out to be negligable.

Using the statistical properties of the interaction matrix elements 
it turns out that a large class
of disordered metallic dots can, under very general conditions, be 
described by a remarkably simple Hamiltonian with only three 
coupling constants, which do not fluctuate.
 In the limit of large  $g_T$, 
the interaction part of the Hamiltonian corresponds to
\begin{equation}  H_{int}=E_c
\hat{N}^2-J(\vec{S})^2+\lambda_{BCS}\hat{T}^{\dagger}\hat{T},
\label{Eq1}
\end{equation} (terms linear in $\hat{N}$ are allowed, but they can 
be included into the one-particle
part of the Hamiltonian) where
$\hat{N}$ is the number operator, $\vec S$ is the spin operator and
$\hat{T}=
\sum_{i} c_{i\uparrow} c_{i \downarrow}$ ($c_{i\uparrow}$ annihilates 
an electron in the $i^{th}$ single
electron orbital with spin $\uparrow$).  In the simple model with 
 short-range interaction and
preserved time reversal ($\beta =1$) the above coupling constants 
have the following form:
\begin{equation}  E_c=\frac{1}{2}\lambda\delta_1; \quad 
J=2\lambda\delta_1;\quad
\lambda_{BCS}=\lambda\delta_1.
\label{couplings}
\end{equation}  If time invariance is broken ($\beta=2$),
$\lambda_{BCS}=0$ since the operator
$\hat{T}$ is incompatible with the symmetry.
 Note that the interaction Hamiltonian Eq.~(\ref{v.int})
represents a particular model, and the expressions for the coupling
constants Eq.~(\ref{couplings}) are valid only for this model.  At the
same time, the effective interaction Hamiltonian,  Eq.~(\ref{Eq1}), is more
robust and depends only on the symmetries of the problem (for instance, on the
absence of the spin-orbit scattering) and on the condition
$g_T\gg 1$.

The first two terms in Eq.~(\ref{Eq1}) represent the dependence of 
the energy of the dot on the
total number of the electrons and on the total spin respectively. 
They commute with each other and
with the single-particle part of the Hamiltonian. 
Therefore, all states of the grain can be classified by $N$ and $S$.
The term proportional to
$\lambda_{BCS}$ appears 
only in the orthogonal case $(\beta = 1)$. 
Provided that $\lambda_{BCS} < 0$ this term leads to the superconducting instability. 
Superconducting correlations are suppressed by the
magnetic field, and thus do not exist for $\beta=2$.

Thus, the general form of the Hamiltonian describing electrons in a 
chaotic  dot which is not
superconducting is given by
\begin{equation} H= \sum_{i} \varepsilon_{i} n_{i} +  E_c N^2 - J 
S(S+1) + g \mu_b S B.
\label{hamil}
\end{equation} 
Note that the only random component of the problem is  the 
single-particle spectrum $\varepsilon_i$, while the exchange
$J$ and charging energy $E_c$ do not fluctuate. 
It should be realized, though, that 
Eqs.~(\ref{Eq1},\ref{hamil}) become exact 
only in the limit $g_T\to\infty$.  
The corrections  to Eqs.~(\ref{Eq1},\ref{hamil}), 
which appear at finite $g_T$ are sometimes important.  
However, if $g_T\gg 1$, these corrections 
do not bring essentially new physics 
to the problem of spin magnetization.

The conductance is calculated using the  many-particle energies $E_N$ 
and  wave functions obtained
numerically  for GOE and GUE random matrix realizations. An example 
of the peak positions and peak
spacing  evolution as functions  of the magnetic field for a 
particular GOE realization is presented
in Fig.\ \ref{fig1} for  (a) $J=0$ and (b) $J=0.1 \delta_1$. For both 
cases we present $\mu_N'=\mu_N -
E_c$. The noninteracting case, Fig.\ \ref{fig1}a, shows  all the 
previously described features of
the Pauli behavior. Once a weak exchange interaction is included  the 
behavior changes
qualitatively, not just at high field, but also in the vicinity of 
$B=0$. In particular, peak
positions are not always paired, resulting in occasional 
field-independent peak spacings.  This
occurs when two consecutive orbitals  are first filled with down spin 
electrons and only later they
acquire  up electrons.  Generally the enhancement  of the spin of the 
dot by $S$ will be accompanied
by a $2S$ bunch peaks moving with the same slope. If the peak spacing
$\Delta_2^N$ is plotted,  two sets of $2S-1$ flat curves sandwiching 
a sloped one will appear.
Changes to the crossing  also attributed to a sudden change of the GS 
spin which is not associated
with a single electron orbital crossing.

As illustrated above, the appearance of  higher spin clearly 
manifests itself  in the peak position
and peak spacing trajectories. Using Eq.(\ref{hamil})  one can 
predict the frequency of spontaneous
magnetization appearances.  For weak exchange,
$J \ll \delta_1$, the probability that  the dot GS  has a spin $S$ is 
determined by the probability
of finding $2S$  orbitals so close to each other that the  gain in 
exchange energy (due to the
polarization) overwhelms the loss in the kinetic energy (due to 
single rather than double
occupations of the orbitals).

Thus the probability for a certain value of ground state magnetization
$S$ boils down to the question of the probability of finding $2S$ 
close single electron levels. That
probability may be estimated using random matrix theory, as we now present.

  The probability  of finding a set of single electron orbital at 
energies $\varepsilon_i$ is given by
Eq. (\ref{repl}). Thus as long as
\begin{equation}
\varepsilon_{i+2S}-\varepsilon_i < J S(S+1) + g \mu_B B S,
\label{spac}
\end{equation} a ground state spin of at least $S$ will appear. 
Taking into account the probability
of Eq. (\ref{repl}), it is possible to write the probability of 
obtaining a ground state spin $S$ as
a function of the linear combination:
\begin{equation} X={{J}\over{\delta_1}} + {{g \mu_B B} \over 
{\lfloor S+{{3}\over{2}} \rfloor
\delta_1}},
\label{xx}
\end{equation} resulting in the following probability
\begin{equation} P_{J,B}(S) = C^{\beta}_S X^{(\beta S + 
1)(2S-1)}(1-K_S^{\beta}X^2),
\label{ph}
\end{equation} The coefficients $C_S^{\beta}$ and $K_S^{\beta}$ 
depend on both $\beta $ and $S$.
Their values for $S = 1, 3/2$  are presented in table I.

\begin{table}[bth] {\protect \narrowtext
\begin{tabular}{|c|c|c|c|c|}
     & $\beta=1$\span\omit &$\beta=2$\span\omit\\ \hline
     & $S=1$&$S=3/2$&$S=1$&$S=3/2$\\ \hline
$C$  & $\pi^2/3$&$9 \pi^4/ 50$&$8\pi^2/9$&$81\pi^6/400$\\
$K$  & $\pi^2/5$&$18 \pi^2/49$&$8\pi^2/25$&$792\pi^2/1225$\\
\end{tabular}
\smallskip\caption{\small The factors $C_S^{\beta}$ and $K_S^{\beta}$ 
appearing in Eq.~(\ref{ph}).}}
\label{tab1}
\end{table}

Thus the RMT model presents a striking conclusion. The influence of 
exchange interactions and
externally applied magnetic field  on the ground state magnetization 
of a quantum dot may be summed
up in a single parameter scaling function, where the  scaling 
parameter is a simple linear
combination of $J$ and $B$. As has been shown from numerical 
simulations\cite{kurland01}, the
scaling holds for larger values of $X$ than expected from the first 
and second term perturbative
analysis presented in Eq. (\ref{ph}).

One may wonder how well does the above picture hold for systems for  which real
correlation between the electron exist, and the dimensionless conductance is not
too large. A canonical example for such systems is the Hubbard model:
\begin{eqnarray} 
\nonumber H= \sum_{i,\sigma} \epsilon_{i} a_{i,\sigma}^{\dag} 
a_{i,\sigma} -t\sum_{<i,j>,\sigma} (a_{i,\sigma}^{\dag} a_{j,\sigma} + h.c) \\
+U \sum_{i} {a_{i, \uparrow}^{\dag} a_{i, \uparrow} a_{i, \downarrow}^{\dag} a_{i,
\downarrow}} + g \mu_B S B
\label{hubbard}
\end{eqnarray} 
where $<i,j>$ denotes nearest neighbor lattice site,
$a_{i,\sigma}^{\dag}$ is an creation operator of an electron at site 
$i$ with spin $\sigma$,
$\epsilon_{i}$ is the site energy, chosen  randomly between $-W/2$  and $W/2$
with uniform
probability, and $U$ is the interaction constant.

The Hilbert space is large even for relatively small systems. For 
example for a $4 \times 4$ lattice
with $6$ electrons for the $S=0$ sector has the size of
$313600$. Using the Lanczos method we obtain the  many-particle 
eigenvalues $E_N(S)$ as function of
the ground state in each spin sector
$S$ for different values of interaction $U$. The disorder was chosen 
as $W=8t$, which corresponds to
the metallic regime, although the value of the dimensionless 
conductance is quite low
$g_T \sim O(1)$.

\begin{figure}[bth]\centering
\includegraphics[width=2.3in]{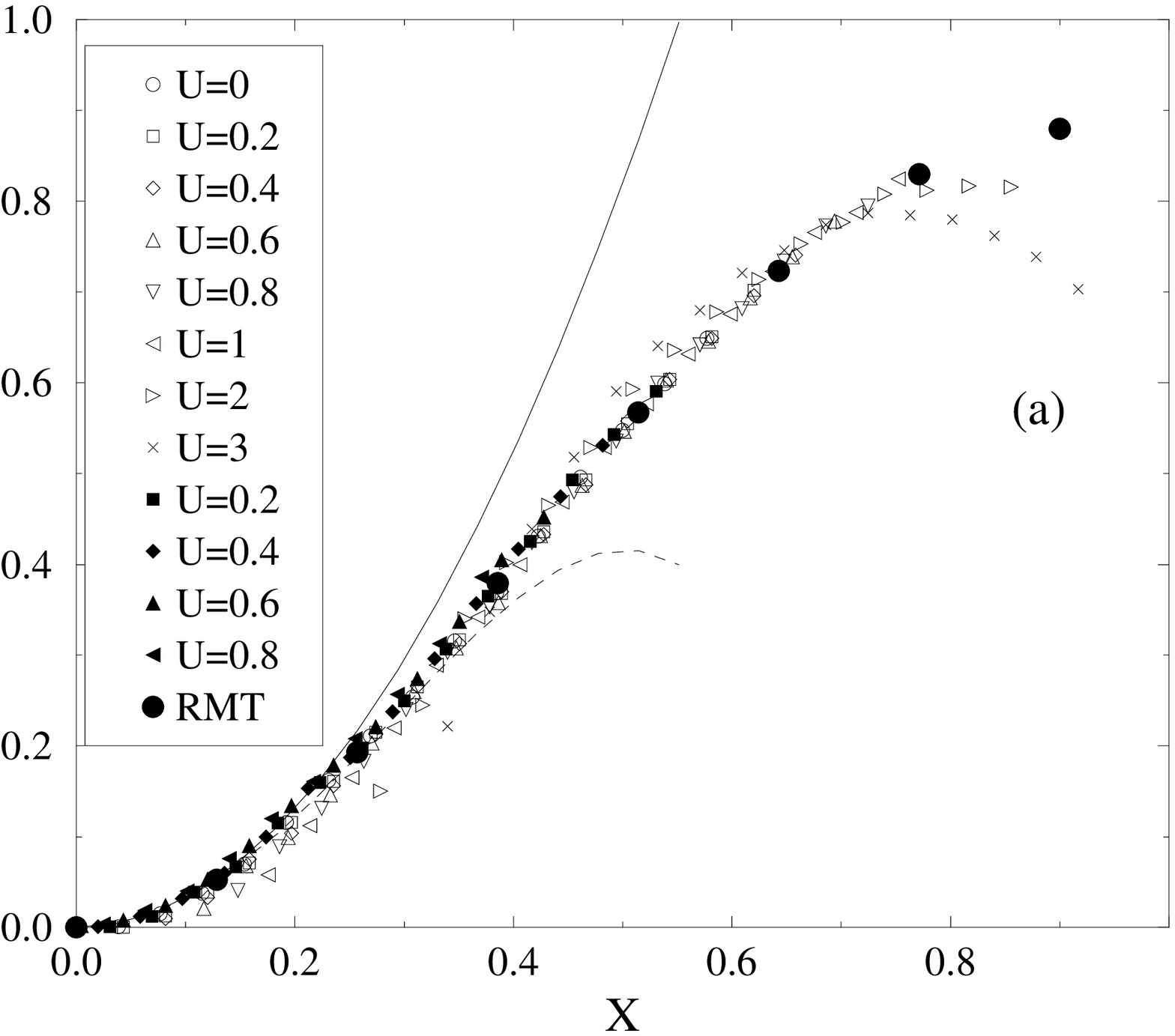}
\includegraphics[width=2.3in]{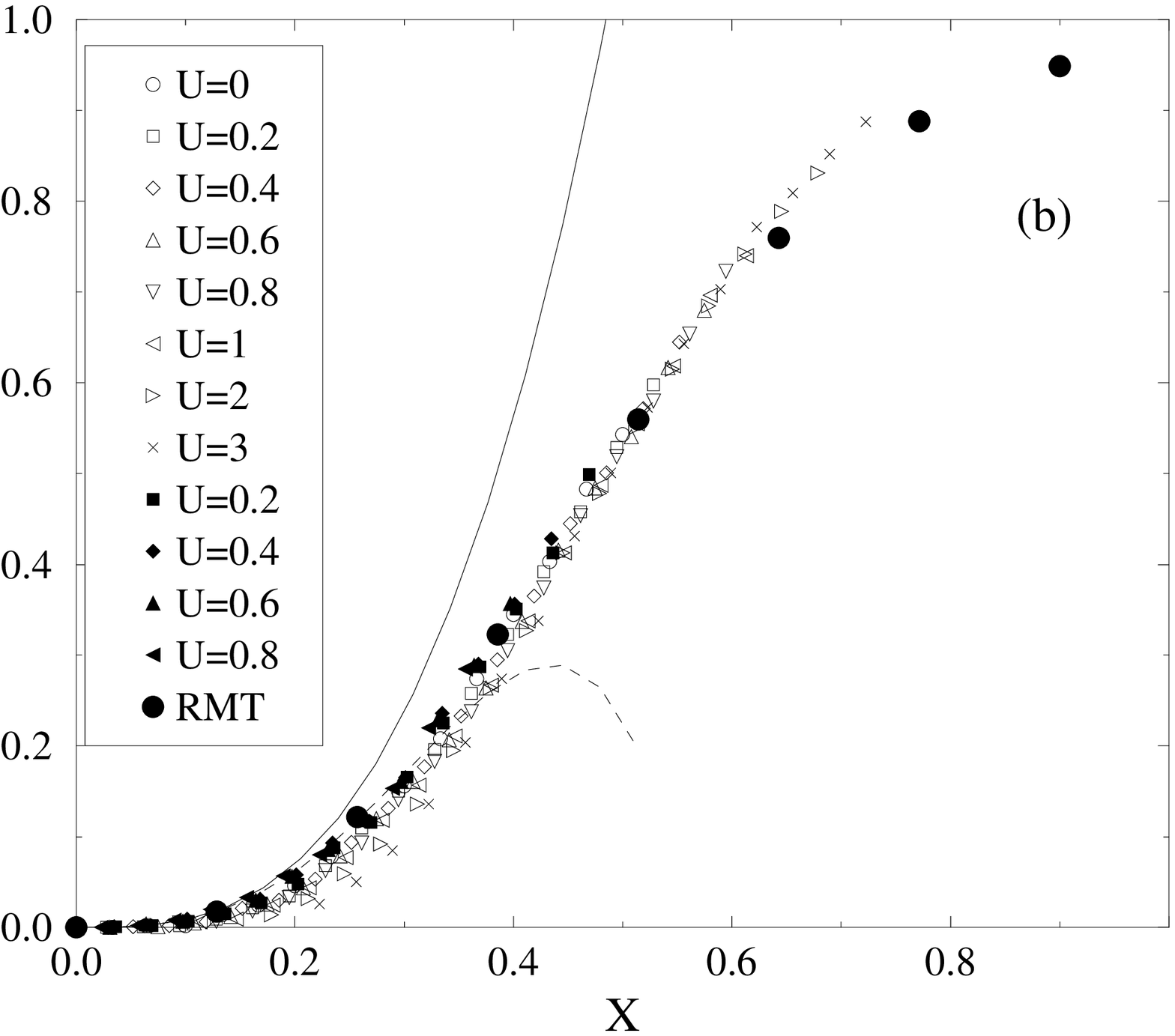}

{\bigskip\caption[]{\small Probabilities of
$S=1$ for $J=0.1\delta_1$  as functions of the magnetic field $B$ for 
(a) the GOE and (b) GUE cases.
The small symbols represent  numerical results for the Hubbard model, 
the large filled symbols
represent numerical results for the RMT model, while the curves 
(solid - first order,  dashed -
second order) represent Eq. (\ref{ph}). }
\label{fig2}}\end{figure}

\begin{figure}[bth]\centering
\includegraphics[width=2.5in]{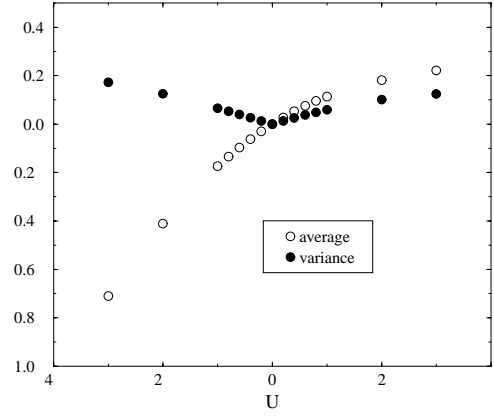}
{\bigskip\caption[]{\small  The 
exchange parameter
$J$  as functions $U$ for a $4 \times 4$ Hubbard model with 6 electrons. }
\label{fig3}}\end{figure}

We calculated the probability for the appearance of a specific value 
of GS spin $S$ for different
values of $U$ and magnetic field $B$. In order to use the scaling 
parameter $X$ we need to deduce the
appropriate  value of $J$ for each value of $U$. This has been done 
by fitting the dependence of the
average lowest energy in a given spin sector
$\langle E(S) \rangle$ to
\begin{eqnarray}
\langle E(S) \rangle = \delta_1 s^2 -JS(S+1).
\label{es}
\end{eqnarray} The distribution
$P_{J(U),B}(S)$ based on numerical results for $1000$ different 
realizations of the Hubbard model,
for different values of $U$ and $B$ for
$\beta=1$ and $\beta=2$ is presented in Fig. \ref{fig2}.  Although 
some deviations do appear,
especially for the higher values of $U$, one can see that the overall 
form of the scaling function
holds remarkably well. Thus, though the Hamiltonian in Eq.\ (\ref{hamil}) is 
obtained under the conditions of large $g_T$ and no correlations in the system, it 
nonetheless appears to  capture much of the physics even for 
a moderate $g_T$  and in the presence of  electron-electron
correlations.  The reason for this is that although neither $J$ nor $E_c$ is
constant once $g_T \sim 1$, their fluctuations remain  small relative to their
average, as is demonstrated in Fig. \ref{fig3}.

\section{Experiment}
\label{s3}

In this section we  describe recent experimental measurements of GS spin for a 
gate-defined GaAs quantum dot containing
roughly 400 electrons. As discussed in the Introduction,  the dot is 
coupled to electron reservoirs
via tunnelling leads (i.e., $g <2e^2/h$ for both leads) so transport 
is dominated by CB effects.
Measurements were carried out at sufficiently low temperature and 
bias that the differences in GS
energies between dots with
$N+1$ and
$N$ electrons should be extractable from CB peak positions, $V_g^{(N)}$.

The dot is formed
at the interface of a GaAs/AlGaAs  heterostructure
($90\,nm$ below the wafer surface) by electrostatic depletion using 
surface gates. A Si delta doping region is located
$40\,nm$ above the heterointerface.  The two-dimensional electron gas 
(2DEG) has density $\sim
2.0\times 10^{11}\, cm^{-2}$ and bulk mobility
$\sim1.4\times10^5\,  cm^2/Vs$, yielding a transport mean free path
$\sim1.5\, \mu m$.  The small dot area, $A\sim 0.25\, \mu m^2$, makes 
transport predominantly
ballistic within the device.  Characteristic energy  scales for the 
measured device include the mean level spacing,
$\Delta = 2\pi\hbar^2/m^*A \sim 30\,\mu eV$, the charging energy, 
$E_C\sim400\mu eV$, and the  Thouless
energy $E_{th} = \hbar v_FA^{-1/2}\sim 340\mu eV$. Measurements were 
carried out in a dilution
refrigerator with a mixing chamber temperature  of  $25\, mK$ using 
standard ac lock-in techniques
with a source-drain bias voltage of $2\,
\mu eV$.  A base electron temperature of  $T_e\sim 50\,mK$  was 
determined from CB peak widths.

To allow the magnetic field to couple predominantly to spin, the 
sample was oriented with the plane of the electron gas along the
axis of the primary solenoid, aligned manually to within 0.5 
degrees.  In addition, a small pair of coils
attached to the vacuum can of the fridge, oriented perpendicular to 
the plane of the sample (see Fig.\ 4), was used  to null out any
perpendicular field from misalignment as well as to explicitly break 
time-reversal symmetry.  Both  primary and trimming coils
were under computer control, allowing sweeps of strictly parallel 
field by trimming out any perpendicular component for any applied 
parallel field.
(We estimate the uncertainty in
$B_\perp$ to be less than $\phi_o/4$ through the dot at 
$B_{\parallel} = 5\,T$.)  Despite this
precise field trimming capability, similar measurements \cite{folk00} 
in larger dots,
fabricated on the same wafer, indicate orbital coupling due to the 
strictly parallel field for
$B_{\parallel}\gtrsim 0.5T$ (evident, for instance, from the
disappearance of the weak localization
feature at $B_\perp = 0$ at higher
$B_\parallel$).  The origin of this surprising coupling---which makes 
interpretation of our high-field
data difficult---is still under investigation.
\begin{figure}[bth]\centering
\includegraphics[width=2in]{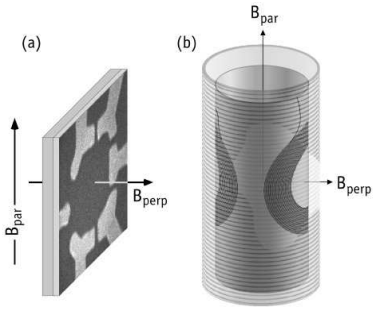}
{\bigskip\caption[]{\small  a) 
Schematic diagram of a gate-defined quantum dot, indicating the
orientations of
$B_\parallel$ and
$B_\perp$ relative to the planar quantum dot. (Orientation of 
$B_\parallel$ within the plane is not
accurately depicted and has not been investigated.)  b)  Diagram 
showing placement of superconducting
coils used to generate $B_\perp$ on vacuum can, inside primary 
solenoid used to generate $B_\parallel$. }
\label{fig4}}\end{figure}

\begin{figure}[bth]\centering
\includegraphics[width=2.75in]{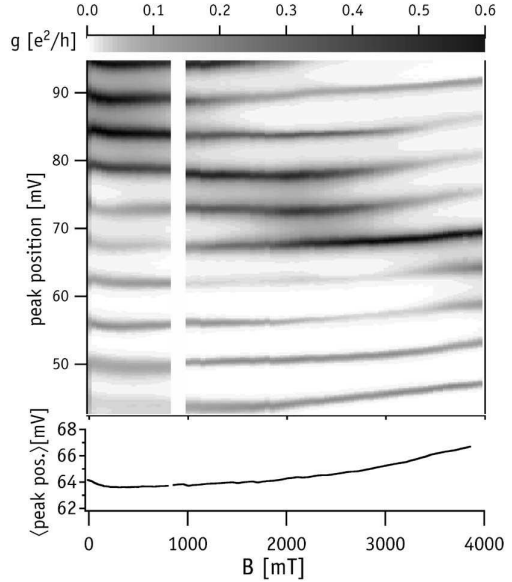}
{\caption[]{\small a) Nine 
consecutive Coulomb blockade peaks measured as a function of
gate voltage $V_g$ and parallel field $B$.  Conductance is shown in 
grayscale, with black
indicating high conductance and white low conductance.  The field 
shown on the bottom axis is the field
strictly parallel to the heterointerface; perpendicular field is held 
constant at $20mT$ as discussed in the
text.  Note that conductance in the valleys is low but not zero, due 
to strong tunnelling in the leads. b)
Average peak position of the peaks in a), showing a diamagnetic shift to higher
energies common to all peaks.  This average is subtracted from all 
peak positions before
further  analysis.}
\label{fig5}}\end{figure}

\begin{figure}[bth]\centering
\includegraphics[width=3in]{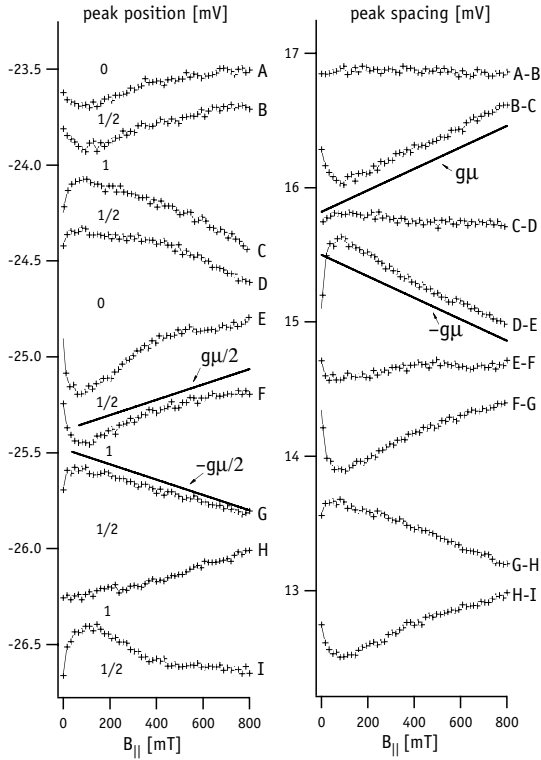}
{\bigskip\caption[]{\small  a) Peak 
positions (with average motion subtracted) as a function of
parallel field for the nine peaks shown in Fig.\ 5a.  Straight lines 
indicate expected peak motion for spin-$\frac{1}{2}$
transitions using the $g$ factor for bulk $GaAs$, $|g| =0.44$. 
Positive slopes indicate transitions to lower
spin states; negative slopes indicate transitions to higher spin 
states.  Numbers indicate a possible
ground-state spin structure for ten consecutive ground states.  Peak 
position data are offset for clarity.  b)
Peak spacings  for the  position data shown in a), offset for 
clarity. Solid lines indicate expected motion of peak spacing for
spin-$\frac{1}{2}$ transitions, using $|g| =0.44$ }
\label{fig6}}\end{figure}

Conductance measurements across ten consecutive Coulomb blockade 
peaks, measured as a function of
$V_g$ and  $B_\parallel$ (i.e., {\em strictly}
$B_\parallel$, properly trimmed), are shown in Fig.\ \ref{fig5}a. 
More positive gate voltage
corresponds to higher energy,  and can be calibrated from the CB 
``diamonds" using high source-drain bias
measurements \cite{dotreview}.  All data are taken with $B_\perp = 20mT$ in order 
to  ensure that time-reversal symmetry is broken, which changes the 
statistics
of dot wave functions.  

In addition to  the individual motions of 
each peak, there is a diamagnetic shift of all peaks (see
Fig.\ \ref{fig5}b), presumably due to the effect of the parallel field on the effective 
well confinement  potential. The slight paramagnetic shift visible 
at low field,
$(B_\parallel<0.2T)$, is not understood at present. In the
following analysis of data, the common curve (in Fig.\ \ref{fig5}b) has been
subtracted from each peak  position.  

Peak positions and peak spacings extracted from the peaks in Fig.\ 5 
are shown in Fig.\ 6.  The slope
of peak positions as a function of
$B_\parallel$ is consistent with a Zeeman energy term $E_S = \pm 
\frac{1}{2} g\mu_B B$, using the
$g$-factor for bulk $GaAs$, $|g| =0.44$.  As discussed in the 
Introduction, alternating slopes for
consecutive peaks would indicate an alternating 
$0,\frac{1}{2},0,\frac{1}{2}...$ GS spin structure.
Our data, on the other hand, shows three consecutive pairs of peaks 
moving with the {\em same}
slope, suggesting the presence of higher spin states. Proposed values 
for the eight consecutive GS
spin states shown here are included in Fig.~6a. We emphasize, 
however, that these are only
plausible values for the spin; it is not possible to determine 
unambiguously the absolute magnitude
of GS spin from measurements of peak position, which reflect {\em 
changes} in spin from the N to N+1
ground states. The  values shown in Fig.\ 6a minimize the ground 
state spin for the system.

\begin{figure}[bth]\centering
\includegraphics[width=2.2in]{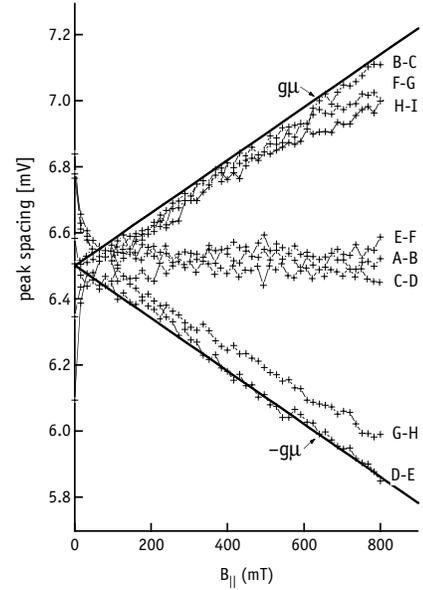}
{\bigskip\caption[]{\small  Peak 
spacings from Fig.\ 6b, offset to
align spacings at $B_\parallel=0$.  Slopes cluster in three branches, 
interpreted as a change from
decreasing to increasing spin transitions (positive slope), two 
consecutive spin transitions in the same
direction (zero slope), and a change from increasing to decreasing 
spin transitions (negative slope). Solid lines indicate expected
motion of peak spacing for spin-$\frac{1}{2}$ transitions, using $|g| 
=0.44$, with no adjustable parameters. }
\label{fig7}}\end{figure}
\begin{figure}[h]\centering
\includegraphics[width=2in]{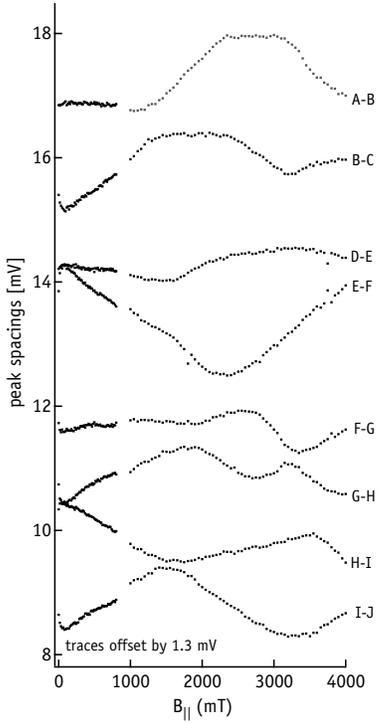}
{\bigskip\caption[]{\small    Peak 
spacings from Fig.\ 6b, shown over an extended range of parallel field, up to
$B_\parallel = 4T$.  Slopes tend to change abruptly between 
linear segments, with rounding between
the segments possibly providing a measure of spin-orbit or other 
spin-mixing interaction. }
\label{fig8}}\end{figure}
\begin{figure}[h]\centering
\includegraphics[width=2.2in]{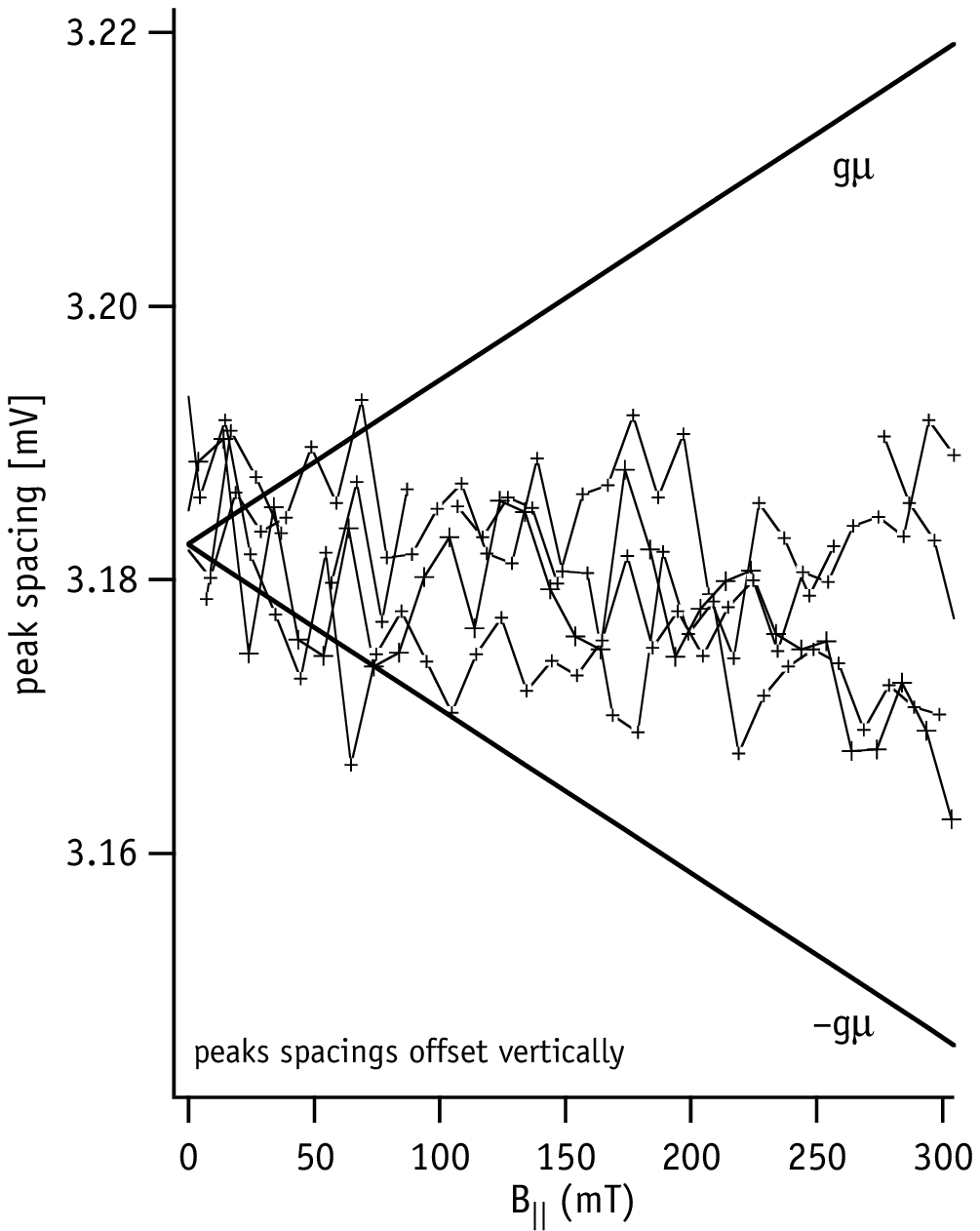}
{\bigskip
\caption[]{\small  Peak spacings for five peaks at low parallel fields, 
measured after
reducing the transmission of the point contacts to allow only weak 
tunnelling between the dot and the leads.
For the four consecutive spacings shown, all have slopes considerably 
less than $g \mu$. }
\label{fig9}}\end{figure}

In the proposed spin labelling scheme, three out of the five even-$N$ 
states have $S=1$, i.e.,
$P(S=1)\sim 0.6$. This fraction of $S=1$ relative to the number of 
$S=0$ states is well beyond the expected value given
reasonable estimates of $J$ for $GaAs$ dots. Of course, with only five 
spin states considered, statistics are
quite poor. Further experiments are needed to see if this discrepancy 
is significant.

Figure 7 shows that peak spacings clearly separate into three 
branches, a top branch with slope
roughly $g\mu$ (corresponding to a GS spin decrement followed by an 
increment) a bottom branch with
slope roughly $-g\mu$ (corresponding to a GS spin increment followed 
by a decrement) and a middle
branch with slope  near zero (corresponding to two consecutive 
increments or decrements).  The
existence of the middle branch is the signature of higher GS spins. 
The good agreement between the
slopes of the upper and lower branches and the expected slopes of 
$g\mu$, as well as the absence of a
range of intermediate slopes suggest that the peak spacing reflects 
spin rather than orbital
coupling. At higher fields, the directions of peak motion change, 
often abruptly and from one
straight segment to another, as seen in Fig.\ 8. This behavior is 
qualitatively similar to the
numerical data in Fig.\ 1b. The rounding of straight segments where 
the slope changes  presumably
results from spin-orbit interaction which mixes spins, and may 
provide a direct measure
of spin-orbit interactions in dots.   It is interesting to note that 
the peaks analyzed in Figs.\ 6
through 8 were measured in a regime of high tunneling conduction in 
the leads. This can be seen by
noting the grayscale of Fig.\ 5.   When the dot is more pinched off 
from the reservoirs, so that the
CB peaks have a height of $0.1\; e^2/h$ or less, peak motion is more 
difficult to interpret, and does
not seem to follow the clear patterns illustrated, for instance, in 
Fig.\ 8. As seen in Fig.\ 9, peak
spacings for four consecutive weak-tunneling peaks at do 
not move with slopes of $g\mu$ at low $B_\parallel$.
Though the data are noisier in this configuration, it is clear that 
the slopes are in all cases less
than $g\mu$. We do not have an explanation for the different behavior 
depending on if the dot is more
closed off or nearly open.

\section{Discussion}
\label{s4}

In conclusion, we have shown that a relatively weak exchange 
interaction qualitatively explains the
deviations from the Pauli picture seen in recent experiments
\cite{rokhinson00,duncan00,folk00}. We calculate the probability that 
different values
of spin appear as a function of exchange interaction and magnetic 
field coupling to the spin.  The strength of
the exchange interaction, $J$, and the effective $g$ factor are the 
only adjustable parameters that
determine the probabilities for a GS of the dot to have any given 
spin at any given magnetic field
for both GOE and GUE cases. In particular we predict that these 
probabilities will follow a one-parameter scaling  law over a wide range of magnetic
fields and  exchange interaction strengths. We also report
preliminary experiments to investigate the exchange effects in the 
ground state spin of quantum dots.  The signature of interaction
effects is the appearance of higher-spin ground states, which show up 
in the experimental data as peak spacing traces that have
zero slope as a function of parallel magnetic field. Such features 
are indeed observed, as seen for instance in Fig.\ 8. Further
experiments are needed to obtain sufficient statistics to 
quantitatively test the predictions of theory.

Support from ARO-MURI DAAG55-98-1-0270 at Princeton and Harvard is gratefully
acknowledged. We  thank B. I. Halperin and L. P.
Rokhinson for many useful discussions, and S. M. Cronenwett for 
assistance with experiments and  analysis. The measured device was
fabricated by S. R. Patel on material grown by C. I. Duru\"oz in the 
lab of J. S. Harris, Jr. at Stanford University. JAF acknowledges 
support as a DoD Fellow.

\end{document}